\documentclass[aps,showpacs,preprint,footinbib,preprintnumbers]{revtex4}

\def\ltap{\ \raise.3ex\hbox{$<$\kern-.75em\lower1ex\hbox{$\sim$}}\ }
\def\gtap{\ \raise.3ex\hbox{$>$\kern-.75em\lower1ex\hbox{$\sim$}}\ }
\def\gl{\ \raise.4ex\hbox{$>$\kern-.75em\lower1ex\hbox{$<$}}\ }

\usepackage[normalem]{ulem}  
\usepackage[dvips]{color} 
\usepackage{amsmath,amssymb}
\usepackage{booktabs}
\usepackage{mathrsfs}
\usepackage[dvipdfmx]{graphicx}
\usepackage{graphicx}
\usepackage{multirow}

\renewcommand\sout{\bgroup \color{red} \ULdepth=-.5ex \ULset}

\newcommand{\Ex}[2]{\ifmmode{#1\times10^{#2}}\else{$#1\times10^{#2}$}\fi}

\def\ket#1{\mathinner{|{#1}\rangle}}

\begin{document}

\title{Decays and productions via bottomonium for $Z_b$ resonances and other $B\bar{B}$ molecules}
\author{S.~Ohkoda$^1$}
\author{Y.~Yamaguchi$^1$}
\author{S.~Yasui$^2$}
\author{A.~Hosaka$^1$}
\affiliation{$^1$Research Center for Nuclear Physics (RCNP), 
Osaka University, Ibaraki, Osaka, 567-0047, Japan}
\affiliation{$^2$KEK Theory Center, Institute of Particle and Nuclear
Studies, High Energy Accelerator Research Organization, 1-1, Oho,
Ibaraki, 305-0801, Japan}

\begin{abstract}
We discuss decays and productions for the possible molecular states formed by 
bottom mesons $B$ ($B^{\ast}$) and $\bar{B}$ ($\bar{B}^{\ast}$).  
The twin resonances found by Belle, $Z_b^{\pm,0}(10610)$ and $Z_b^{\pm}(10650)$, 
are  such candidates.   
The spin wave functions of the molecular states are rearranged into those of 
heavy and light spin degrees of freedom by using the re-coupling formulae of 
angular momentum.  
By applying the heavy quark symmetry we derive model independent relations 
among various decay and production rates, which can be tested in experiments.  
\end{abstract}
\pacs{14.40.Rt, 12.39.Hg, 13.20.Gd}
\maketitle

Exotic hadrons are  important  for the study of hadron physics.
Not only multiquarks, but also
hadronic molecules, provide an alternative opportunity for the study 
of structures of hadrons.
Hadronic molecules are especially interesting because they can appear in the  threshold region of hadron-quark-clusters which saturate 
the color dependent force of QCD and weakly interact each other.   
Thus, they emerge as loosely bound states of hadrons with spatially 
extended property, which should be distinguished from the conventional hadrons 
where the minimal number of constituent quarks are considered to be rather localized.  

One of the good candidates of such a hadronic molecule is a pair of 
$Z_b(10610)$ and $Z_b(10650)$ which has been recently observed 
in the processes $\Upsilon(5S) \rightarrow \pi\pi \Upsilon(nS) (n=1,2,3)$
and $\Upsilon(5S) \rightarrow \pi\pi h_b(kP) (k=1,2)$
by Belle group~\cite{Collaboration:2011gja,Belle:2011aa}.
The reported masses and widths of the two resonances are
$M(\mathrm{Z}_{\mathrm{b}}(10610)) = 10607.2 \pm 2.0$ MeV, 
$\Gamma (\mathrm{Z}_{\mathrm{b}}(10610)) = 18.4 \pm 2.4$ MeV and
$M (\mathrm{Z}_{\mathrm{b}}(10650)) = 10652.2 \pm 1.5$ MeV,
$\Gamma (\mathrm{Z}_{\mathrm{b}}(10650))=11.5 \pm 2.2$ MeV,
which appear very close to the $B\bar{B}^{\ast}$ and $B^{\ast}\bar{B}^{\ast}$ thresholds, respectively, 
and have relatively narrow decay widths.
The facts that (1) the minimal constituent quarks must be four because of isospin one in their observed quantum numbers 
$I^G(J^{PC}) = 1^+(1^{+-})$, 
(2) the mass difference between $Z_b(10610)$ and $Z_b(10650)$ is so small as 40 MeV, and (3) the decays into a bottomonium 
both through heavy spin conserved and non-conserved processes~\cite{Adachi:2011ji} indicate 
that  $Z_b(10610)$ and $Z_b(10650)$ are likely $B^{(\ast)}\bar{B}^{(\ast)}$ molecular states \cite{Bondar:2011ev,Voloshin:2011qa}.
Here, we use a notation $B^{(\ast)}$ to stand for $B$ or $B^{\ast}$.

In our previous work, we have investigated all possible  $B^{(\ast)} \bar{B}^{(\ast)}$ 
bound or resonant states having exotic quantum numbers 
of the total angular momentum $J \leq 2$~\cite{Ohkoda:2011vj}.
To verify whether these theoretically expected states exist or not in experiments,
it is useful to study their production and decay processes.
In this paper, we derive relative rates of each transition  when we consider that $B^{(\ast)}\bar{B}^{(\ast)}$ molecular states obey the heavy quark symmetry.
The heavy quark symmetry allows only the processes where heavy quark spin is conserved,
leading to selection rules among certain classes of transitions.
To derive them, we consider the spin structure of the mesons in terms of spin re-coupling 
formula which is equivalent to Fierz rearrangement.
By rearranging  the two heavy quarks in $B^{(\ast)}$ and $\bar{B}^{(\ast)}$ mesons of a molecular state,
we can separate the heavy quark spin and the spin of light degrees of freedom in heavy quark limit.

In general, the total angular momentum ${\bf J}$ of a hadron  is a conserved quantity.
The spin of heavy quark ${\bf S}_{H}$ is also conserved 
in the limit of heavy quark mass, $m_{Q} \rightarrow \infty$.
Then we can define the spin of light degrees of freedom ${\bf S}_l$ by
\begin{eqnarray}
 {\bf S}_l \equiv {\bf J} - {\bf S}_{H} \,,
\end{eqnarray}
which is also conserved.
Generally speaking, the spin of light degrees of freedom has complex structure.
For instance, a $Q\bar{q}$ meson includes an anti-quark $\bar{q}$, gluons, an arbitrary number of 
 $q\bar{q}$ pairs and angular momentum ${\bf L}$ as the light degrees of freedom.
Although they are not conserved separately, the sum of them is conserved 
in the heavy quark limit.
This quantity is referred to simply as ``light spin'' because it includes all degrees of freedom
except for the heavy quark spin.
Therefore, we can describe the spin structure of a hadron containing heavy quarks 
in terms of the good quantum numbers ${\bf S}_{H}$ and ${\bf S}_{l}$.
As an example, the spin structure of $B$ meson is 
$B(0^-) = (\frac{1}{2}_{H} \otimes \frac{1}{2}_l)|_{J=0}$ and that of $B^{\ast}$ meson is 
$B^{\ast}(1^-) = (\frac{1}{2}_{H} \otimes \frac{1}{2}_l)|_{J=1}$.

Now we consider  $Z_b(10610)$ and $Z_b (10650)$.
We assume that the main component of the wave function of  $Z_b (10610)$ is 
$\frac{1}{\sqrt{2}}(B\bar{B}^{\ast} - B^{\ast}\bar{B})(^3S_1)$ and
that of the $Z_b (10650)$ is $B^{\ast}\bar{B}^{\ast}(^3S_1)$.
Because these masses are close to  $B\bar{B}^{\ast}$ and $B^{\ast}\bar{B}^{\ast}$ thresholds  
respectively, and  the rate of $D$-wave mixing is not large as the previous study indicates
that the probability of  $\frac{1}{\sqrt{2}}(B\bar{B}^{\ast} - B^{\ast}\bar{B})(^3D_1) $ is about $9\%$ and  that of $B^{\ast}\bar{B}^{\ast}(^3D_1)$ is about $6\%$ in the total wave function of $Z_b(10610)$~\cite{Ohkoda:2011vj}.
Let us now employ the spin re-coupling formula with 9-j symbols to analyze the spin structure of 
$\frac{1}{\sqrt{2}}(B\bar{B}^{\ast} - B^{\ast}\bar{B})(^3S_1)$ and 
$B^{\ast}\bar{B}^{\ast}(^3S_1)$.
This standard formula is written as
\begin{eqnarray}
  [[l_1 , s_1]^{j_1} , [l_2 , s_2]^{j_2}]^{J} = \sum_{L,S} 
\hat{j_1} \hat{j_2} \hat{L} \hat{S} 
\begin{Bmatrix}
 l_1 & s_1 & j_1 \\
 l_2 & s_2 & j_2 \\
 L & S & J
\end{Bmatrix}
 [[l_1 , l_2]^{L} , [s_1 , s_2]^{S}]^{J} \, ,
\label{spin_recoupling}
\end{eqnarray}
where $[j_1 ,j_2]^J$ means that the angular momenta $j_1$ and $j_2$ are coupled to the total angular
momentum $J$, and  $\hat{J}=\sqrt{2 J +1}$.
By using this, the heavy and light spins of  $B\bar{B}^{\ast}(^3S_1)$ and $B^{\ast}\bar{B}(^3S_1)$ are re-coupled as
\begin{eqnarray}
\ket{B\bar{B}^{\ast}(^3S_1)} &=&  [[b \bar{q}]^{0} , [\bar{b}
 q]^{1}]^{1} \notag \\ 
&=& \sum_{H,l} \hat{0}\hat{1}\hat{H}\hat{l} 
\begin{Bmatrix}
 1/2 & 1/2 & 0 \\
 1/2 & 1/2 & 1 \\
 H & l & 1
\end{Bmatrix}
\left[ [b\bar{b}]^H, [\bar{q}q]^l \right]^1 \notag \\
&=& \frac{1}{2}\left[ [b\bar{b}]^0, [\bar{q}q]^1 \right]^1 
- \frac{1}{2} \left[ [b\bar{b}]^1, [\bar{q}q]^0 \right]^1
+ \frac{1}{\sqrt{2}}\left[ [b\bar{b}]^1, [\bar{q}q]^1 \right]^1 \notag \\
&=& \frac{1}{2} (0^-_H \otimes 1^-_l) - \frac{1}{2} (1^-_H \otimes
 0^-_l)  + \frac{1}{\sqrt{2}} (1^-_H \otimes 1^-_l) \,,
\label{BB*3S}
\end{eqnarray}
\begin{eqnarray}
 \ket{B^{\ast}\bar{B}(^3S_1)} &=&  [[b  \bar{q}]^{1} , [\bar{b} 
 q]^{1}]^{0}  \notag \\ 
&=& -\frac{1}{2} (0^-_H \otimes 1^-_l) + \frac{1}{2} (1^-_H \otimes
 0^-_l)  + \frac{1}{\sqrt{2}} (1^-_H \otimes 1^-_l) \,,
\label{B*B3S}
\end{eqnarray}
which give the spin structure of $\frac{1}{\sqrt{2}}(B\bar{B}^{\ast} - B^{\ast}\bar{B})(^3S_1)$. 
For $B^{\ast}\bar{B}^{\ast}(^3 S_1)$, we have
\begin{eqnarray}
\ket{B^{\ast}\bar{B}^{\ast}(^3 S_1)}  && =  \left[ [b\bar{q}]^1, [\bar{b}q]^1 \right]^1 \notag \\
 && = \sum_{H,l} \hat{1}\hat{1}\hat{H}\hat{l} 
\begin{Bmatrix}
 1/2 & 1/2 & 1 \\
 1/2 & 1/2 & 1 \\
 H & l & 1
\end{Bmatrix}
\left[ [b\bar{b}]^H, [\bar{q}q]^l \right]^1 \notag \\
&& = \frac{1}{\sqrt{2}}\left[ [b\bar{b}]^0, [\bar{q}q]^1 \right]^1 + 
\frac{1}{\sqrt{2}}\left[ [b\bar{b}]^1, [\bar{q}q]^0 \right]^1 \notag \\
&& = \frac{1}{\sqrt{2}}(0_H^- \otimes 1_l^-) + \frac{1}{\sqrt{2}} (1_H^- \otimes 0_l^-) \, .
\label{B*B*3S}
\end{eqnarray}
If the structure of $Z_b$'s is dominated by $B^{(\ast)}\bar{B}^{\ast}(^3S_1)$,
their spin configurations are given from~(\ref{BB*3S})-(\ref{B*B*3S})  as
\begin{eqnarray}
 \ket{Z_b (10610)} &=& \frac{1}{\sqrt{2}}(0_H^- \otimes 1_l^-) - \frac{1}{\sqrt{2}} (1_H^- \otimes 0_l^-) \label{Zb} \,, \\
\ket{Z_b (10650)} &=& \frac{1}{\sqrt{2}}(0_H^- \otimes 1_l^-) + \frac{1}{\sqrt{2}} (1_H^- \otimes 0_l^-) \label{Zb'} \, .
\end{eqnarray}
It is important to note that $Z_b$'s have the same fraction of a heavy quark spin singlet component and 
a triplet component.
Due to the heavy quark spin symmetry, the former component decays into spin-singlet bottomonium  $h_b (kP)$ 
($\ket{h_b (kP)}  = 0^-_H \otimes 1_l^-$) with a pion emission in $S$-wave.
By contrast, the latter decays into a spin-triplet bottomonium $\Upsilon (nS)$ 
($\ket{\Upsilon (nS)} = 1^-_H \otimes 0^+_l$) with a pion emission in $P$-wave.
Therefore, the rates of the decays into a spin singlet state $Z_b \rightarrow h_b(kP) \pi$  
should be comparable to that of the decays into a spin triplet state  
$Z_b \rightarrow \Upsilon(nS) \pi$.
Hence, the experimental observation of the two pion emission of $\Upsilon(5S)$ can be 
explained if the process occurs through the intermediate state of $Z_b$~\cite{Adachi:2011ji}.
These arguments have been already made by Fierz transformation  in Ref.~\cite{Bondar:2011ev,Voloshin:2011qa}.
Here we have shown the same results in terms of the 
spin re-coupling formula (\ref{spin_recoupling}),
which is also applied to other processes in a straightforward manner.

Next we consider  the neutral resonance $Z_b^0(10610)$ recently observed 
in the processes $\Upsilon(5S) \rightarrow \pi^0 \pi^0 \Upsilon(1S,2S)$ by Belle group~\cite{Adachi:2012im}.
It is possible for  $Z_b^0 (10610)$ to decay into 
$\Upsilon \pi^0$, $h_b \pi^0$, $\eta_b \gamma (\rho^0)$
and $\chi_{bJ}\gamma$ ($J=1,2,3$) from the viewpoint of the conservation of quantum numbers and kinematics.
In general, the decay  $\Upsilon(5S) \rightarrow \eta_b \pi^0 \gamma (\rho^0)$ should be
suppressed because this process requires heavy quark spin flip.
However, going through $Z_b$, the decay into the singlet bottomonium state
is allowed.

$Z_b^0(10610)$ can decay into $\chi_{bJ}$ ($J=1,2,3$) by a photon emission.
The radiative transition into a bottomonium is a new decay pattern for $Z_b$, which cannot be seen in charged $Z_{b}^{\pm}$.
It can be used to investigate the structure and  interaction of  $Z_b$.
Here, we estimate the ratio of decays $Z_b^0 \rightarrow \chi_{bJ}\gamma$.
To do this, we derive the spin structure of $\chi_{bJ}\gamma$.
Since $Z_b^0(10610)$ has $I(J^{PC}) = 1(1^{+-})$ as a possible isospin partner of $Z_b^{\pm}(10610)$, 
the photon of the $\chi_{b0}\gamma$ decay channel must carry  $J^P = 1^+$,  which corresponds 
to an M1 transition.
Therefore, the spin structure of the photon can be written as $\ket{\gamma (M1)} = 0_H^+ \otimes 1_l^+$.
The spin structure of $\chi_{b0}$ is $\ket{\chi_{b0}} = (1^-_H \otimes 1^-_l)|_{J=0}$.
Applying  the re-coupling formula to the spin of $\chi_{b0}$ and  photon, 
we find
\begin{eqnarray}
 \ket{\chi_{b0} \gamma(M1)} &=& (1_H^- \otimes 1_l^-)|_{J=0} \otimes (0_H^+ \otimes 1_l^+) \notag \\
&=& \frac{1}{3}(1_H^- \otimes 0_l^-) -\frac{1}{\sqrt{3}}(1_H^- \otimes 1_l^-)|_{J=1} + \frac{\sqrt{5}}{3} (1_H^- \otimes 2_l^-)|_{J=1} \,. \label{chib0M1}
\end{eqnarray}
The same consideration is applied to the decay  $Z_b^0(10610) \rightarrow \chi_{b1}\gamma$,
where  M1 and  E2 transitions are possible.
The spin structures of these decay channels are given by
\begin{eqnarray}
 \ket{\chi_{b1} \gamma (M1)} \, &=& -\frac{1}{\sqrt{3}}(1_H^- \otimes 0_l^-) 
+\frac{1}{2} (1_H^- \otimes 1_l^-)|_{J=1} +\frac{15}{6}(1_H^- \otimes 2_l^-)|_{J=1} \,, \\
 \ket{\chi_{b1} \gamma (E2)} \, &=& -\frac{1}{2}(1_H^- \otimes 1_l^-)|_{J=1} 
+\frac{\sqrt{3}}{2}(1_H^- \otimes 2_l^-)|_{J=1} \,.
\end{eqnarray}
The decay  $Z_b^0(10610) \rightarrow \chi_{b2}\gamma$ can also occur in 
M1 and E2 transitions and the spin structures of these states are given by
\begin{eqnarray}
\ket{\chi_{b2}\gamma (M1)} \, &=& \frac{\sqrt{5}}{3} (1_H^- \otimes 0_l^-) 
+ \frac{\sqrt{15}}{6}(1_H^- \otimes 1_l^-)|_{J=1} +\frac{1}{6}(1_H^- \otimes 2_l^-)|_{J=1} \,, \\
\ket{\chi_{b2}\gamma (E2)} \, &=& \frac{\sqrt{3}}{2}(1_H^- \otimes 1_l^-)|_{J=1} 
+\frac{1}{2}(1_H^- \otimes 2_l^-)|_{J=2} \,. \label{chib2E2}
\end{eqnarray}
Because $Z_b^0 (10610)$ has the structure (\ref{Zb}), its radiative decay  is  possible only through M1 transitions,
and  (\ref{chib0M1})-(\ref{chib2E2}) imply that 
the decay ratio is given as
\begin{eqnarray}
 \begin{matrix}
\Gamma (Z_b^0 \rightarrow  \chi_{b0} \gamma) &:& \Gamma (Z_b^0 \rightarrow  \chi_{b1} \gamma) 
&:& \Gamma (Z_b^0 \rightarrow  \chi_{b2} \gamma) \\
1 &:& 3 &:& 5 
\end{matrix} \,.
\label{ratio:chigamma}
\end{eqnarray}
Note that  the differences of the phase spaces are neglected,
and furthermore overlaps of the meson wave functions are assumed to be the same,
which is the case
when the spatial wave functions of $\chi_{bJ}$ have the same node quantum number.
Considering the phase space factors  proportional to the cube of the photon energy $\omega_{J}$,
we find the relation for
$Z_b^0(10610) \rightarrow \chi_{bJ}(1P)$
\begin{eqnarray}
 \begin{matrix}
\Gamma (Z_b^0 \rightarrow  \chi_{b0}(1P) \gamma) &:& \Gamma (Z_b^0 \rightarrow  
\chi_{b1}(1P) \gamma) &:& \Gamma (Z_b^0 \rightarrow  \chi_{b2}(1P) \gamma) \\
1 &:& 2.6 &:& 4.1
\end{matrix} \,,
\label{ratio:chi1gamma}
\end{eqnarray}
and for
$Z_b^0(10610) \rightarrow \chi_{bJ}(2P)$
\begin{eqnarray}
 \begin{matrix}
\Gamma (Z_b^0 \rightarrow  \chi_{b0}(2P) \gamma) &:& \Gamma (Z_b^0 \rightarrow  
\chi_{b1}(2P) \gamma) &:& \Gamma (Z_b^0 \rightarrow  \chi_{b2}(2P) \gamma) \\
1 &:& 2.5 &:& 3.8
\end{matrix} \,,
\label{ratio:chi2gamma}
\end{eqnarray}
which can be compared with experimental data.
So far we have assumed that $Z_b(10610)$ wave function is dominated by $^3S_1$ of 
$\frac{1}{\sqrt{2}}(B\bar{B}^{\ast} - B^{\ast}\bar{B})$.
It should be noted that a $^{3}D_{1}$ component exists with a small fraction in the $\frac{1}{\sqrt{2}}(B\bar{B}^{\ast} - B^{\ast}\bar{B})$ molecular state \cite{Ohkoda:2011vj}.
The spin structure of $\frac{1}{\sqrt{2}}(B\bar{B}^{\ast} - B^{\ast}\bar{B})(^3D_1) $ is
$\frac{1}{\sqrt{2}}(0_H^- \otimes 1^-_l) + \frac{1}{\sqrt{2}}(1^-_H \otimes 2^-_l)|_{J=1}$,
which modifies slightly the relations (\ref{ratio:chi1gamma}) and (\ref{ratio:chi2gamma}). 
Moreover, it implies that E2 transition in the processes $Z_b^0 \rightarrow \chi_{bJ}\gamma$ must
 occur mediated by the $D$-wave component.
This is also an interesting point to be studied in experiments.

Now, we consider the production and the decay for recently predicted isotriplet $B^{(\ast)}\bar{B}^{(\ast)}$
molecular states having positive $G$-parity~\cite{Ohkoda:2011vj}.
These states can be produced by one pion emission in $P$-wave from $\Upsilon(5S)$.
We introduce new notations $W_{bJ}^{PC}$ for the lowest state and 
$W_{bJ}^{\prime PC}$ for the first excited state.
The quantum numbers and the main components of these $B^{(\ast)}\bar{B}^{(\ast)}$ molecular states
are 
\begin{eqnarray}
 W^{--}_{b0} &\,  1^+(0^{--})  &: \frac{1}{\sqrt{2}}(B\bar{B}^{\ast} +B^{\ast}\bar{B})(^3P_0) \,, 
 \label{W0}\\
 W^{\prime --}_{b1} &\,  1^+(1^{--}) &: \frac{1}{\sqrt{2}}(B\bar{B}^{\ast} +B^{\ast}\bar{B})(^3P_1) \,, \\
 W^{--}_{b1} &\,  1^+(1^{--})  &: B\bar{B} (^1P_1)  \,, \\
 W^{\prime --}_{b2} &\,  1^+(2^{--}) &: B^{\ast}\bar{B}^{\ast}(^5P_2)  \,, \\
 W^{--}_{b2} &\,  1^+(2^{--}) &: \frac{1}{\sqrt{2}}(B\bar{B}^{\ast} +B^{\ast}\bar{B})(^3P_2)  \,. \label{W2}
\end{eqnarray}
We note that, in principle, $B^{\ast}\bar{B}^{\ast}(^1P_1)$ and $B^{\ast}\bar{B}^{\ast}(^5P_1)$ 
in $I^G(J^{PC}) = 1^+(1^{--})$ states can be mixed.
However, the previous study indicates 
that $W_{b1}^{--}$ and $W_{b1}^{\prime --}$ are close to each threshold
of $B\bar{B}$ and $B\bar{B}^{\ast}$, which means $B^{\ast}\bar{B}^{\ast}$ components are suppressed.
There could be also coupled channels $\frac{1}{\sqrt{2}}(B\bar{B}^{\ast} +B^{\ast}\bar{B})(^3F_2)$ and 
$B^{\ast}\bar{B}^{\ast}(^5F_2)$ components in $1^+(2^{--})$ states.
However,
we expect that  the probabilities of these components are small  due to  high angular momentum.
Therefore, from Eqs.~(\ref{W0})-(\ref{W2}) the corresponding spin structures of $W_{bJ}^{--}$ states are given as
\begin{eqnarray}
W^{--}_{b0} &:&  (1^-_H \otimes 1^+_l)|_{J=0}  \, , \\
W^{\prime --}_{b1} &:& -\frac{1}{\sqrt{3}}(1^-_H \otimes 0^+_l) 
+ \frac{1}{2}(1^-_H \otimes 1^+_l)|_{J=1} +\frac{\sqrt{5}}{2\sqrt{3}}
(1^-_H \otimes 2^+_l)|_{J=1}   \, ,\\
W^{--}_{b1} &:& \frac{1}{2}(0^-_H \otimes 1^+_l) 
+ \frac{1}{2\sqrt{3}}(1^-_H \otimes 0^+_l) -\frac{1}{2}(1^-_H \otimes
1^+_l)|_{J=1} +\frac{\sqrt{5}}{2\sqrt{3}}(1^-_H \otimes 2^+_l)|_{J=1} \, , \\
W^{\prime --}_{b2} &:&  \frac{\sqrt{3}}{2}(1_H^- \otimes 1_l^+)|_{J=2} 
+ \frac{1}{2} ( 1^-_H \otimes 2^+_l)|_{J=2} \, , \\
W^{--}_{b2} &:& -\frac{1}{2}(1^-_H \otimes 1^+_l)|_{J=2}  
+ \frac{\sqrt{3}}{2} ( 1^-_H \otimes 2^+_l)|_{J=2} \, .
\end{eqnarray}
Remarkably, the heavy quark spin singlet state exists only in $W_{b1}^{--}$.
Therefore decays of $W_{bJ}^{--}$ into  
a heavy quark spin singlet state of bottomonium and light hadrons are forbidden 
except for $W_{b1}^{--}$, although their quantum numbers and  kinematics allow them.
Only $W_{b1}^{--}$ can decay into $h_b \pi$ or $\eta_b \rho(\gamma)$.
In contrast, all  $W_{bJ}^{--}$ can decay into a heavy quark spin triplet state, e.g.
$\Upsilon \pi$ in $P$-wave ($\ket{\Upsilon \pi}_{P{\mbox -wave}} = (1^-_H \otimes 1^+_l)|_{J=0,1,2}$).
The decay ratio of the processes $W_{bJ}^{--} \rightarrow \Upsilon \pi$ is obtained as
\begin{eqnarray}
 \begin{matrix}
 \Gamma (W^{--}_{b0} \rightarrow \Upsilon \pi) &:& \Gamma (W^{\prime --}_{b1} \rightarrow \Upsilon \pi) &:& \Gamma (W^{--}_{b1} \rightarrow \Upsilon \pi)
 &:&  \Gamma (W^{\prime --}_{b2} \rightarrow \Upsilon \pi)
 &:& \Gamma (W^{--}_{b2} \rightarrow \Upsilon \pi) \\
4 &:& 1 &:& 1 &:& 3 &:& 1
\end{matrix} \,.
\end{eqnarray}
The decay processes $\Upsilon(5S) \rightarrow W_{bJ}^{--}\pi \rightarrow \Upsilon(nS)\pi\pi$ are not yet observed, 
though they are similar to $\Upsilon(5S) \rightarrow Z_b \pi \rightarrow \Upsilon(nS)\pi\pi$.
However we expect that the production rates of $W_{bJ}^{--}$ is small compared with 
that of $Z_b$, since the $W_{bJ}^{--}$ transition is mediated by a pion emission in $P$-wave.
High statistics and refined analysis of experiments will establish
the presence or absence of $W_{bJ}^{--}$ and their production and decay properties. 
Finally, the production ratio of $W_{bJ}^{--}$ from $\Upsilon(5S)$ is obtained as
\begin{eqnarray}
 \begin{matrix}
  f(W_{b0}^{--} \pi) &:& f(W_{b1}^{\prime --} \pi) &:& f(W_{b1}^{--} \pi) &:& 
f(W_{b2}^{\prime --} \pi) &:& f(W_{b2}^{--} \pi)  \\
 2 &:& 9 &:& 4.5 &:& 9 &:& 12 
 \end{matrix} \,,
\end{eqnarray}
where we find the production rate of $W_{b2}^{--}$ is favored,
while the production of $W_{b0}^{--}$ is suppressed.

In summary, we have derived the model independent relations among various decay and production rates
for possible $B^{(*)} \bar B^{(*)}$ molecular states under the heavy quark symmetry.  
Part of decay properties of $Z_b$(10610) and $Z_b$(10650) are well explained
and the possible decay patterns for neutral $Z_b^0$(10610) are discussed in the present framework.
We have shown that the $W_{bJ}^{--}$ decay 
into a spin singlet bottomonium is forbidden except for $W_{b1}^{--}$.  
We have also predicted the production rate of various $W_{bJ}^{--}$ 
through the one pion emission of $\Upsilon(5S)$.
All of them can be tested experimentally and will provide important information to 
further understand the exotic structure of the new particles.  


\section*{Acknoledgements}

This work is partly supported by the Grant-in-Aid for
Scientific Research on Priority Areas titled ``Elucidation of
New Hadrons with a Variety of Flavors'' (E01: 21105006).


\begin{thebibliography}{99}

\bibitem{Collaboration:2011gja} 
  I.~Adachi [Belle Collaboration],
  arXiv:1105.4583 [hep-ex].

\bibitem{Belle:2011aa} 
  A.~Bondar {\it et al.}  [Belle Collaboration],
  Phys.\ Rev.\ Lett.\  {\bf 108}, 122001 (2012)
  [arXiv:1110.2251 [hep-ex]].

\bibitem{Adachi:2011ji} 
  I.~Adachi {\it et al.}  [Belle Collaboration],
  Phys.\ Rev.\ Lett.\  {\bf 108}, 032001 (2012)
  [arXiv:1103.3419 [hep-ex]].

\bibitem{Adachi:2012im} 
  I.~Adachi {\it et al.}  [Belle Collaboration],
  arXiv:1207.4345 [hep-ex].

\bibitem{Bondar:2011ev}
  A.~E.~Bondar, A.~Garmash, A.~I.~Milstein, R.~Mizuk and M.~B.~Voloshin,
  Phys.\ Rev.\  D {\bf 84}, 054010 (2011)
  [arXiv:1105.4473 [hep-ph]].

\bibitem{Voloshin:2011qa} 
  M.~B.~Voloshin,
  Phys.\ Rev.\ D {\bf 84}, 031502 (2011)
  [arXiv:1105.5829 [hep-ph]].

\bibitem{Ohkoda:2011vj} 
  S.~Ohkoda, Y.~Yamaguchi, S.~Yasui, K.~Sudoh and A.~Hosaka,
  Phys.\ Rev.\ D {\bf 86}, 014004 (2012)
  [arXiv:1111.2921 [hep-ph]].


\end{thebibliography}
\end{document}